# Towards smart optical focusing: Deep learning-empowered wavefront shaping in nonstationary scattering media

YUNQI LUO[†1], SUXIA YAN[†1], HUANHAO LI[†2], PUXIANG LAI[*2], AND YUANJIN ZHENG[*1]

[1]*School of Electrical and Electronic Engineering, Nanyang Technological University, Singapore, 639798*
[2]*Department of Biomedical Engineering, Hong Kong Polytechnic University, Hong Kong SAR, China*
[†] *These authors contributed equally to this work.*
[*] *Corresponding authors: puxiang.lai@polyu.edu.hk; yjzheng@ntu.edu.sg;*

**Abstract:** Optical focusing at depths in tissue is the Holy Grail of biomedical optics that may bring revolutionary advancement to the field. Wavefront shaping is a widely accepted approach to solve this problem, but most implementations thus far have only operated with stationary media which, however, are scarcely existent in practice. In this article, we propose to apply a deep convolutional neural network named as ReFocusing-Optical-Transformation-Net (RFOTNet), which is a Multi-input Single-output network, to tackle the grand challenge of light focusing in nonstationary scattering media. As known, deep convolutional neural networks are intrinsically powerful to solve inverse scattering problems without complicated computation. Considering the optical speckles of the medium before and after moderate perturbations are correlated, an optical focus can be rapidly recovered based on fine-tuning of pre-trained neural networks, significantly reducing the time and computational cost in refocusing. The feasibility is validated experimentally in this work. The proposed deep learning-empowered wavefront shaping framework has great potentials in facilitating optimal optical focusing and imaging in deep and dynamic tissue.

## Introduction

When light enters a disordered medium which is thicker than a few scattering mean free path $l$ (~0.1 mm for human skin), light will undergo multiple scattering due to the mismatch of refractive index [1]. After scattering, a lot of information carried by the incident light is lost. Therefore, scattering has impeded high-resolution optical delivery and imaging through or within thick scattering media, such as biological tissues. If light is coherent, scattered light along different optical paths interfere randomly, forming speckles, whose intensity distribution can be recorded outside the medium using cameras. Although it seems that speckles are randomly formed and distributed, the way light is scattered is actually deterministic within a certain time window (usually referred as speckle correlation time)[2]. Built upon this property, various approaches have been inspired, such as time reversal[1,3-8], wavefront shaping[2,9-13], and memory effect[1,14-17], to obtain optical focusing and imaging through scattering media. Time reversal methods, such as TRUE[3] and TROVE[5], take advantage of guide stars (e.g. focused ultrasonic modulation) to encode diffused light, then only the encoded light is time-reversed and focused inside the scattering medium. Wavefront shaping modulates the phases of light incident into the scattering medium based on the measurement of transmission matrix[10,12,13] or the maximization of feedback provided by the optical[9] or photoacoustic signal strength[2], with a goal to pre-compensate for the scattering-induced phase distortions. Representative algorithms to obtain the optimal compensation include continuous sequential algorithm (CSA)[18], genetic algorithm (GA)[19], or phase-conjugation of the transmission matrix[10]. As for the memory effect method, image information is encoded in the autocorrelation of the measured speckles as long as the imaging area is within the memory effect range, so that images can be reconstructed from speckles with iterative phase retrieval algorithms[1,15].

Each of the aforementioned three approaches has its own advantages and limitations. For example, wavefront shaping is attractive due to its plain working principle and experimental setup. Approaches to sense and control wavefront have already been reported a lot[20-22]. However, it is inherently time consuming as many iterations are required regardless of the optimization algorithm[23]. For this reason, almost all wavefront shaping implementations reported thus far have operated with stationary medium such as diffusers. Nevertheless, it is hardly possible to find perfectly stationary media in reality. Even for seemingly stable objects, for example, diffusers, tissue-mimicking phantoms, and multimode fibers, their properties or states might be altered due to inevitable environment disturbance. When media are perturbed, the focusing will be degraded or even disappear completely. To recover the focusing, the wavefront shaping iterations have to be repeated from the beginning each time the scattering medium changes, which is again a tedious process[24]. Memory effect can deal with circumstances with slight perturbations, but when the correlation function of two electric fields drops below 0.5, memory effect vanishes[25]. This problem impedes the implementation of wavefront shaping to more general and realistic applications.

Deep learning, often known as deep neural networks (DNNs), are promising for their superior ability in revealing complex relationships through transforming representations at one level to a higher and more abstract level[26]. So far, DNNs have been widely used to solve inverse problems such as denoising[27-29], deconvolution[30-32], image reconstruction[33-38], and super-resolution imaging[39-41]. The idea has also been exploited to focus light[42] and reconstruct images[43-45] through static scattering media. For example, Turpin *et al* introduced neural networks for binary amplitude modulation, and focused light through a single diffuser[42]; Li *et al*. trained U-Net with speckles generated by various objects through different diffusers[43]. The pre-trained network was capable to be generalized to "unseen" objects or diffusers. All these diffusers, however, are of the same macroscopic parameters. In this paper, we take one step further. For the first time to the best of our knowledge, we introduce a DNN framework to tackle the challenge of optical refocusing in nonstationary scattering media, which is of great significance in practice. It is known that media before and after changes are correlated to some extent, and the existence of correlation has been exploited for imaging through scattering media[15,46-48] (See "Methods" for speckles correlation theory in random media).Taking advantage of such correlation, DNNs can adapt to different environments through fine tuning, and fast recovery of a focal point from perturbations may be feasible. For example, Sun *et al*.[49] trained five different neural network models to reconstruct blurred images by first classifying scattered images then feeding the images into one of the five pre-trained models to recover them. Note that, however, considering the computation time and memory budget, it is difficult and impractical to train hundreds of neural network models to cover all possible scattering conditions; using five models probably only gets a rough classification and reconstruction. In this work, we do not depend on any classification or pre-trained models. When a scattering medium is changed due to environmental perturbations, we use a small amount of newly available samples to fine tune the neural network, so that it can specifically and precisely model the changed medium. The hypothesis is strongly supported by experimental results, suggesting that the DNNs have great potentials in facilitating smart optical optimal focusing and imaging in deep tissues. The memory and time cost in computation for wavefront shaping optimization and re-optimization can be significantly reduced. As known, optical fields at depths *in vivo* alter at the speed of milliseconds[50], posing high demand on processing time, which has not been effectively resolved yet in the field. The fast optical focusing recovery capability enabled by DNNs opens up a new path to settle this challenge, which will benefit a wide range of biomedical imaging, sensing, control and treatment applications in deep biological tissue.

**Results**

As illustrated in Fig. 1(a), when coherent light with different phase patterns propagates through a disordered medium, light will be scattered, and speckles will be formed and recorded outside the medium by a camera. In order to resolve inverse scattering problems which are nonlinear and ill-posed, iterative optimization algorithms with regulation are required[51]. So far almost all state-of-art iterative reconstruction algorithms are the cascades of convolutional operations and nonlinear pointwise operations[39], which are similar with the structure of conventional convolutional neural networks (CNNs). This property suggests that CNNs are intrinsically suited to tackle wavefront shaping problems. In addition, deep neural networks are particularly powerful in solving inverse problems[52] and forming images with noisy, uncertain, or ill-posed measurements[34]. Thanks to these merits, in this study, we directly apply a multi-input deep CNN named as ReFocusing-Optical-Transformation-Net (RFOTNet) to model the inverse scattering process which can be simplified as, establish the relationship between the speckle intensity distribution and its corresponding incident optical wavefront, and more importantly, to fast recover the focal point when the medium is perturbed based on the speckle correlation theory in random media. The process can be formulated as a supervised learning problem[53] (See "Methods" for detailed theoretical analysis). Considering that spatial light modulators (SLMs) are often used to modulate incident optical wavefront in practice, as shown in Fig. 1(a), in this article, we employ SLM patterns to safely represent the incident optical phase patterns.

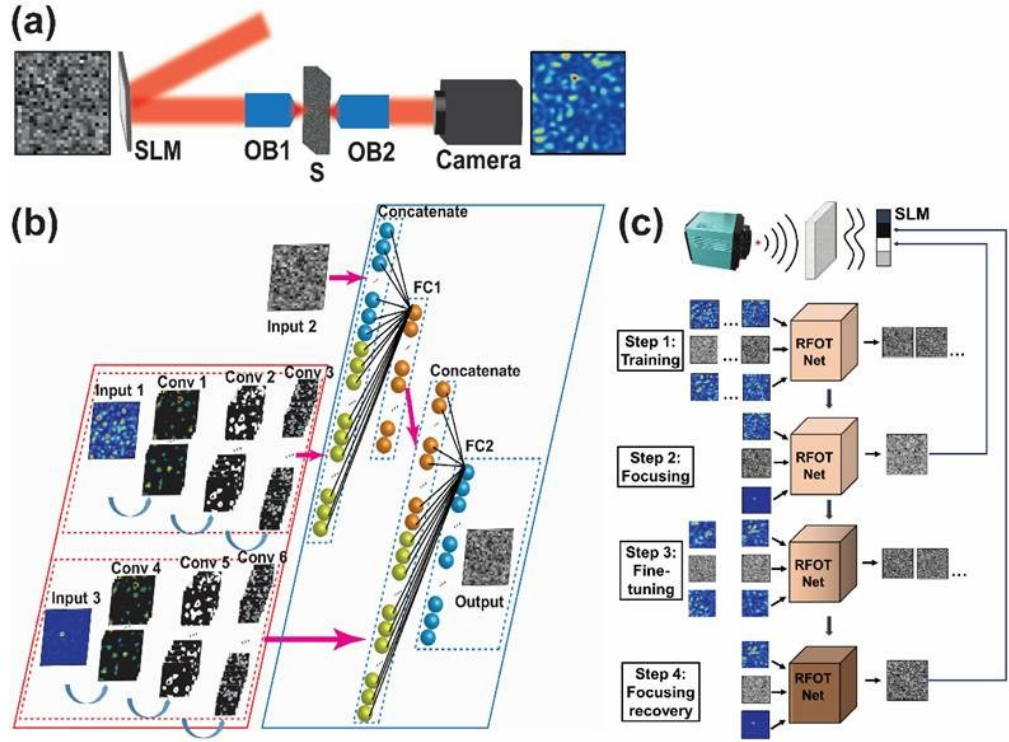

Fig 1. (a) A brief illustration of light scattering. The phase patterns of light are modulated by a spatial light modulator (SLM), then light is focused onto a scattering layer (S) by an objective lens (OB1). Scattered light transmitted the scattering layer is collected by another objective lens (OB2) and recorded by a camera. (b) Structure of the proposed RFOTNet. RFOTNet has three inputs and one output. Input 1 is the speckle pattern, while the corresponding SLM pattern is noted as Input 2. Input 3 is the targeted speckle pattern, while RFOTNet output is the SLM pattern needed in order to get Input 3 through the scattering medium. (c) Illustration of the working principle of the proposed method. First of all, samples are collected to train RFOTNet. After training, the RFOTNet can model the inverse scattering function, and output a SLM pattern that can focus light through the scattering medium. When the medium is altered due to perturbations, the focal point can be degraded or even lost. Benefiting from the correlation of the medium speckles before and after the change, only a small amount of extra samples are sufficient to fine-tune the pre-trained RFOTNet in step 3. After the tuning, the RFOTNet can adapt to the concurrent medium state and recover the optical focus.

The structure of the RFOTNet is shown in Fig. 1(b), which is specifically designed to achieve light focusing and refocusing through nonstationary scattering media (See "Methods" for detailed RFOTNet structure). The RFOTNet has three inputs and one output. They work collaboratively as a team to learn the information of the medium, establishing the transformation from speckles to their corresponding SLM patterns. Comparing with a conventional CNN which simply learns the mapping from speckles to SLM patterns, the RFOTNet demonstrated superior performance in terms of convergence speed, light focusing and refocusing results (see "Supplementary information" for comparison). The results indicate that with the introduction of multiple inputs, the RFOTNet learns richer information about the scattering processes so that it can establish a more accurate inverse scattering model. More importantly, since optical fields of the medium before and after change are correlated, during the fine-tuning of RFOTNet, the information of the medium before perturbation is retained to some extent. Meanwhile, RFOTNet learns the differences between the media, and adjusts the weights to adapt to the new medium, contributing greatly to increase fine-tuning speed and accuracy, which has been proved in Supplementary information. Fig. 1(c) illustrates the working principle of the proposed method. In the first step, samples are collected for neural network training. After training, RFOTNet will precisely model the inverse scattering function and establish the relationship between SLM patterns and the corresponding speckles patterns. Then a focused (speckle) pattern is sent to the RFOTNet through Input 3 as the target, and RFOTNet predicts the SLM pattern required in order to generate the focused target in the current system. Light modulated by this SLM pattern propagates through the same scattering medium, and is supposed to focus to a single point, which should be the same as the target. Affected by unavoidable perturbations from the environment, the scattering medium is changed to some extent, causing the focus to be degraded or even lost. To recover this focus, samples collected in the changed medium are used to fine-tune the RFOTNet trained in Step 1. As the optical fields of the medium before and after the change are correlated, the amount of samples for fine-tuning is much less than that used for retraining a network. After the directed adjustment, the RFOTNet will be able to adaptively model the perturbed medium, and produce an SLM pattern that can recover the optical focusing. The scattering medium can be perturbed continuously, and each time it is altered, the RFOTNet is tuned to adapt to the concurrent state, and help to refocus light. These four steps function together, assuring optimal light focusing in the nonstationary medium.

To verify the feasibility of the proposed method, simulation was conducted. We used a transmission matrix generated from circularly symmetric Gaussian distribution[19] to represent light scattering paths inside a disordered medium. The size of the SLM patterns was $32 \times 32$, while the size of speckle patterns was $64 \times 64$. In Step 1, 10,000 samples were generated based on this transmission matrix for RFOTNet training. After training, a SLM pattern can be predicted by the RFOTNet. With such a SLM pattern, light was successfully focused through the scattering medium as shown in Fig. 2(a). Note that Fig. 2a-c use the same colormap. Next, we modelled different levels of perturbations by adding matrices with different variances to the original transmission matrix. The variances are also of circularly symmetric Gaussian distribution. This procedure can be regarded as the original transmission matrix being affected by independent circular symmetric Gaussian noises with different real-valued variances. After perturbation, the optical focus was degraded (Fig. 2b). To quantitatively evaluate the focusing performance, the enhancement $\eta$ is defined as the ratio between the optimized intensity at the chosen position and the average intensity before optimization[9]. We then introduce correlation coefficient of speckles to identify the extent of medium change; two speckle patterns were recorded before and after the medium change, while the SLM pattern remained identical. As shown in Fig. 2(d), with increased variance, which suggests larger perturbations, both the correlation coefficient and enhancement $\eta$ drops accordingly. Relative $\eta$ is calculated by the ratio of $\eta$ after medium change over the original $\eta$. New samples were generated from the altered transmission matrix to fine-tune the pre-trained RFOTNet. The simulated focusing recovery effect is shown in Fig. 2(c).

We generated twenty different matrices to model different levels of perturbations, and for each matrix, simulation was conducted five times. Results were averaged and shown in Fig. 2(e), which illustrates the amount of fine-tuning samples required to recover the original focusing under different levels of medium change. When the correlation coefficient is around 0.6, the amount of fine-tuning samples is only a quarter of that used for training RFOTNet in Step 1, while the original enhancement ratio can be recovered. When the correlation coefficient drops to 0.3 or even lower, the amount of fine-tuning samples is larger than the half required in Step 1. However, the amount of fine-tuning samples increases sharply when the correlation coefficient drops below 0.2, which means little information of the original status is preserved. These simulation results are consistent with predictions based on the correlation theory[54]. With the increase of perturbations, correlation coefficient drops following a single-sided Gaussian distribution[25,55], which is reflected in Fig. 2(d). When the perturbation is mild, as shown in Fig. 2(e), the correlation between the speckles is governed by the short-range correlation $C^{(1)}_{aba'b'}$, but when the perturbation becomes stronger, the correlation may fall into the long-range correlation $C^{(2)}_{aba'b'}$, whose magnitude is much smaller than $C^{(1)}_{aba'b'}$ (See "Methods" for details). In this case, the distinction due to the change is enlarged, and hence the fine-tuning of the pre-trained neural networks needs more samples and time to adapt itself to model the new situation. Nevertheless, so long as the correlation coefficient remains to be larger than 0.2, the amount of samples for fine-tuning is convincingly smaller than the original number (10,000 samples).

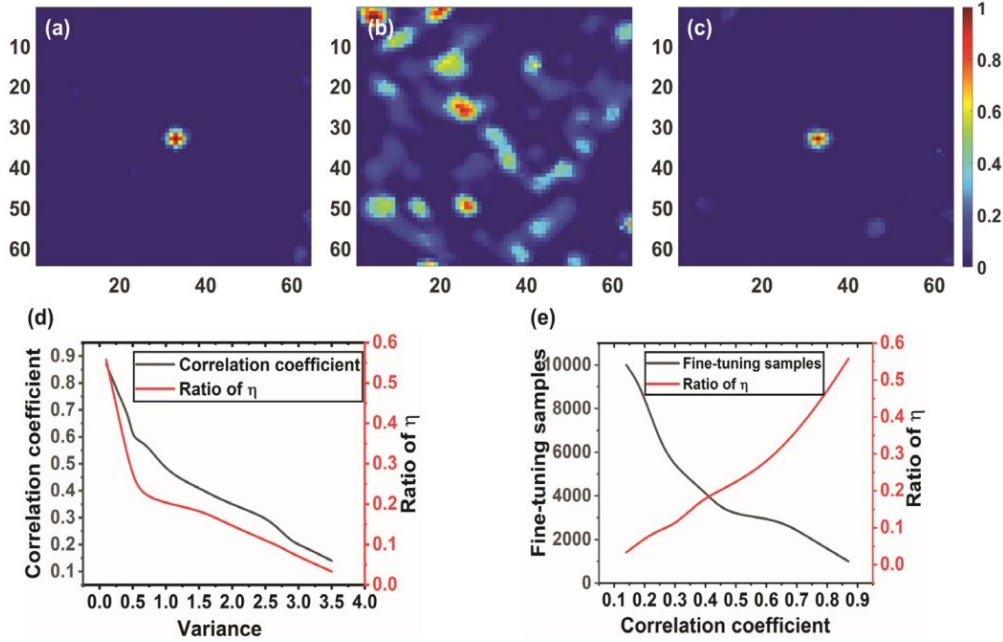

Fig. 2. Simulation results. (a) Focused speckle pattern obtained after Step 1. (b) Speckle pattern after the medium change. (a) and (b) are recorded with identical modulation pattern on the SLM. (c) A new focused speckle pattern can be obtained after fine-tuning the pre-trained RFOTNet. (a), (b), and (c) use the same colormap. (d) Different levels of variances/perturbations result in different levels of transmission matrix change. (e) The relationship between the degree of medium change and the amount of required samples for fine-tuning to recover the focusing.

The experimental setup is illustrated in Fig. 3. The resolution of the SLM screen is 1280 × 1024, and it is divided to 32 × 32 macropixels to display the SLM patterns, i.e., one macropixel contains 40 × 32 pixels. The size of speckle patterns recorded by the camera is 64 × 64 pixels. In experiment, we use 32 grey levels in SLM to represent phase values from 0 to $2\pi$. In experiment, for Step 1, 10,000 pairs of samples were collected for training, and the epoch was set to 10. The reason why 10,000 samples were used here is that, based on the training results with different amounts of samples, training RFOTNet with 10,000 samples could balance the trade-off between focusing performance and training cost including time and memory. Before training, the intensity of all collected speckles and SLM patterns are respectively normalized to between 0 and 1. Training time was approximately 2 minutes. After training, a targeted speckle (shown as Fig. 4(m)) with a single speckle grain was sent to the RFOTNet, predicting an output, which was then loaded onto the SLM to modify the incoming optical wavefront. The final speckle pattern recorded using the predicted SLM pattern was shown in Fig. 4(a). As seen, light was focused to a single point with diameter ~30 μm, and the enhancement $\eta$ was 64. After training, the RFOTNet has learned the information of the current medium state, and is able to focus light through the diffuser. It is worth noting that although here we only focus light to a single position, the trained RFOTNet is capable to focus light to an arbitrary position on the image plane.

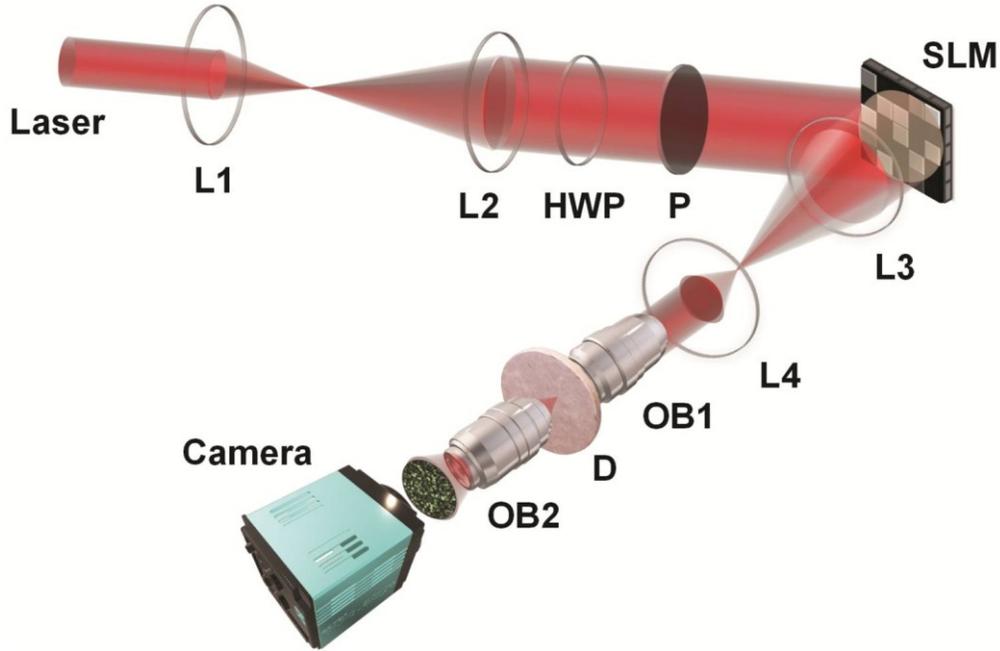

Fig.3. Schematic of the experimental setup. Light is expanded by two lenses (L1 and L2), and then half-waveplate (HW) and polarizer (P) adjust the polarization state of light incident onto the spatial light modulator (SLM). Light is modulated and reflected by the SLM, then passes through two lenses (L3 and L4) and is focused onto a diffuser (D) surface by an objective lens (OB1). Scattered light is collected by another objective lens (OB2) and recorded by a camera.

Then perturbations were applied to the scattering medium by mechanically moving the diffuser along one direction. After the movement, the speckle pattern also changed, as shown in Fig. 4(b). Note that Fig. 4a-l use the same colormap and scale. While the SLM pattern remained identical, the focusing performance was degraded as indicated by the appearance of more speckle grains in the background. The correlation coefficient between (a) and (b) is 0.5221, with which the enhancement $\eta$ dropped from 64 to 27 (Fig. 4n, Group 1). To recover the focal point, 3,000 pairs of SLM patterns and speckles were collected to fine-tune the pre-trained RFOTNet. The reason why 3,000 samples was chosen is that, based on simulation results, when the correlation coefficient is around 0.5, ~3,000 samples is required to recover the original focusing performance. Iterating 60 epochs took about 3 minutes. The target pattern was the same as the one used in Step 1. The SLM pattern predicted by the fine-tuned RFOTNet resulted in a new focus shown in Fig. 4(c), whose enhancement $\eta$ is much higher than Fig. 4(b) as the background is significantly suppressed.

After Group 1 demonstration, experiment conditions were manually varied, and the whole experiment was repeated three more times from Steps 1 to 4. The amount of samples in pre-training and fine-tuning, the RFOTNet structure, and the training settings were all identical to the first experiment (Group 1). Results are illustrated in Fig. 4 as Groups 2-4, respectively. As given in Fig. 4(n), the focused speckle patterns had enhancements $\eta$ of 49, 36, and 23, respectively, after Step 1. After the medium (diffuser) was moved laterally by various distances, the focusing performance was obviously degraded (Figs. 4e, h, and k). More and more bright speckle grains appeared in the background, and eventually the original focal point was submerged and hard to be distinguished (Fig. 4k). In these three groups, the correlation coefficients with the original patterns were 0.4822, 0.4252, and 0.6976, respectively, and the enhancement $\eta$ dropped to 22, 17, and 10, respectively. Then for each group, 3,000 samples were used to fine-tune each individual pre-trained RFOTNet. Focusing recovery results are shown in Figs. 4f, i, and l, with the enhancements $\eta$ improved to 40, 32, and 22, respectively. There is no doubt that after fine-tuning the RFOTNet, the clear focus is recovered from medium perturbations.

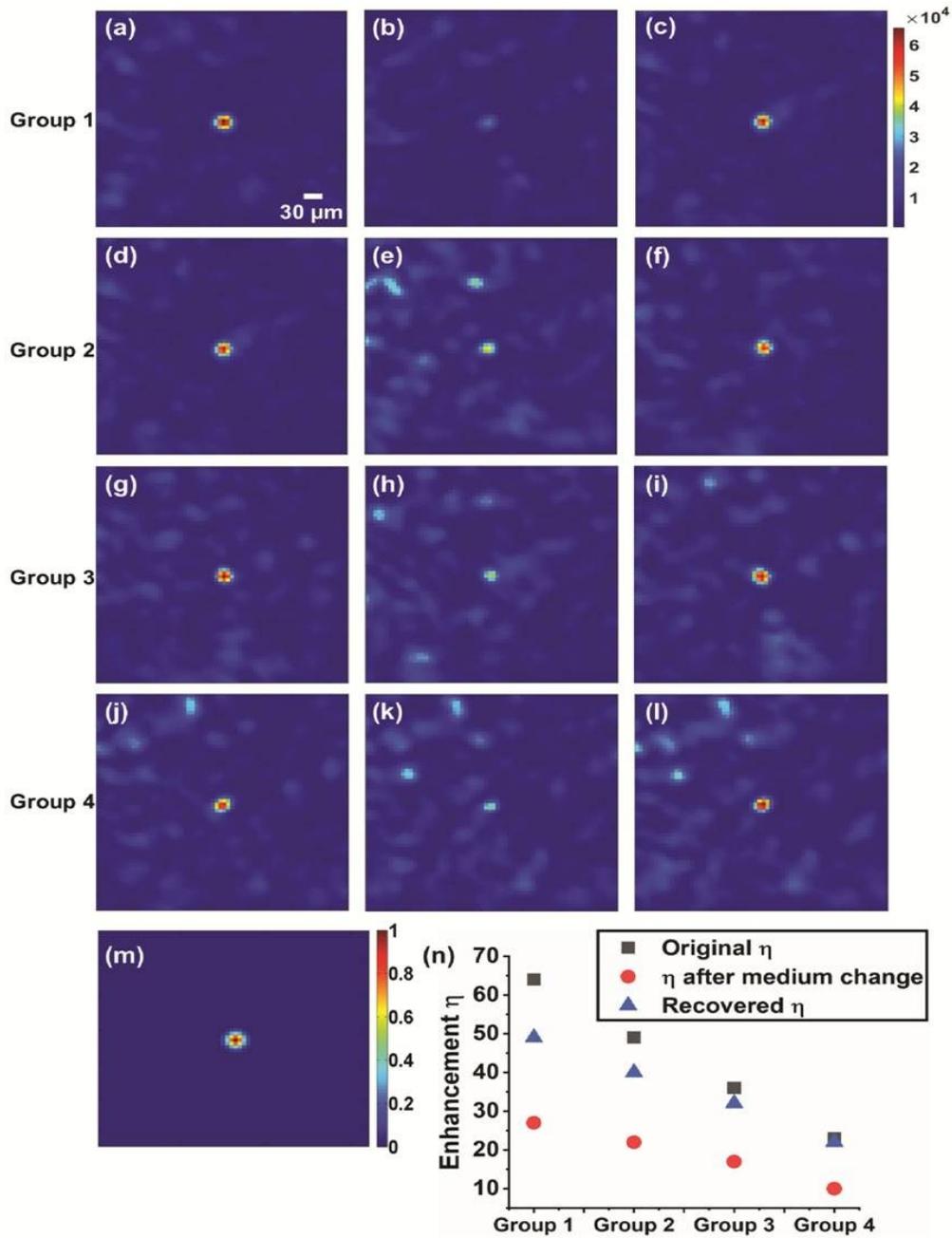

Fig 4. Experimental results of four different groups to achieve focusing recovery through the diffuser. (a)-(c) Experimental results of Group 1. (a) is the original focal pattern obtained using the RFOTNet trained in Step 1, (b) is the pattern after the perturbation, and (c) is the recovered focal pattern after fine-tuning the RFOTNet. (d)-(f), (g)-(i), and (j)-(l) are the speckle patterns corresponding to before change, after change, and after RFOTNet fine-tuning for Groups 2 to 4, respectively. (a)- (l) use the same colormap and scale. (m) The target speckle. (n) Enhancement $\eta$ of the focus of speckle patterns. Black squares indicate the $\eta$ after Step 1, red circles indicate the $\eta$ after the medium change, and blue triangles indicate the $\eta$ of recovered focus after Step 4. All figures from (a) to (l) are of the same scalar bar.

It is worth noting that the correlation coefficients before and after medium change in this study were all larger than 0.2, for which one may doubt whether it was within the traditional memory effect range[56].

However, retaining the original phase modulation pattern on the SLM can no longer keep light focused after the medium change in experiment, which is consistent with the finding reported by Benjamin et al.[25]. As known, if it is within the memory effect range, when the SLM pattern is shifted slightly, say, along a lateral direction, the resultant focal point shifts laterally accordingly, although the focal intensity decreases with the displacement distance, following a bell-shaped curve[25]. In our experiment, the maximal medium movement is smaller than 2.5 μm, which is much less than the individual element size of the camera (6.5 μm) and the SLM (12.5 μm). So, the movement of the focal point is not obvious. But when the scattering medium is disturbed while the SLM pattern remains unchanged, the intensities of the focal point are significantly reduced, resulting in much more random speckle patterns that may even submerge the focal point. It is thus confirmed that using traditional memory effect methods cannot keep the clear focal point after the medium change.

It also should be highlighted that if conventional iterative wavefront shaping[9] or transmission matrix methods[10] are applied to solve this problem, as discussed earlier, the whole optimization process has to be repeated from the beginning. Imagine the SLM pattern is 32 × 32 as in our experiment, 4,096 measurements are required to calculate the transmission matrix. Moreover, interference between the modulated light and a reference light is required, which significantly increases the system complexity and reduces the efficiency of the SLM considering that part of the SLM pixels work as the reference. As for the iterative optimization methods, such as CSA, the phase value of each macropixel on SLM has to be cycled from 0 to $2\pi$ in multiple steps. As in our experiment, $2\pi$ is divided into 32 grey levels, which requires altogether 32,768 measurements to calculate the phase mask towards an optimum focusing. In comparison, using the proposed method in this article, only 3,000 (*one order of magnitude less*) measurements are used to recover the focal point from disturbance. In addition, CSA optimizes each macropixel independently, hence the detected intensity improvement at the output plane is small, making it susceptible to noises[57]. In contrast, our approach optimizes all the SLM pixels together, leading to significant increase in signal-to-noise ratio and computational efficiency.

**Discussion**

Optical focusing plays a central role in many biomedical applications, which, however, is challenging at depths in tissue due to the strong scattering of light. In recent years, progresses in wavefront shaping have been achieved to focus light tightly and efficiently through or within thick scattering media. Most of these works (including CNN-based studies)[2,9,13,42], however, have been limited to mechanically stationary samples as the each optimization process is specified to only one state of the medium. When the medium is perturbed or starts to move, the quality of light focusing will degrade in terms of focal intensity and focal-to-background ratio. If the change is beyond the memory effect regime, with existing methods, a new time- and/or resource-demanding optimization process is required to recover the focusing. To tackle such a challenge, in this study, for the first time to the best of our knowledge, we propose to fine-tune the pre-trained RFOTNet so as to recover the degraded or even extinctive optical focus after perturbation much more rapidly than the conventional methods.

In order to manifest the performance of our method against conventional memory effect more clearly, experiments were conducted and results are shown below. Fig. 5(a) is a focused pattern achieved using RFOTNet. Then the diffuser was moved left by 1μm, resulting in obvious degradation in focusing, as shown in Fig. 5(b). The correlation between (a) and (b) is 0.5303, while the enhancement $\eta$ in (b) drops to only 32% of (a). The correlation between the original focused speckle (Fig. 5a) and speckles after shifting the diffuser is a function of translation distance, as shown in Fig. 5(d). The focal intensity also drops as the translation distance increases, following a bell shape[25] as illustrated in Fig. 5(e). Relative intensity is calculated against the focal intensity in (a). Despite that 1 μm does not exceed the memory effect range, when the SLM pattern was shifted left accordingly, it resulted in a speckle pattern shown in Fig. 5(c), failing to recover the focal point. Therefore, slight disturbance in the medium can lead to severe decrease in speckle correlation as well as focal intensity, while traditional memory effect is no longer able to achieve a clear focal point. With our method, as shown above, a clear focused speckle pattern can be reliably recovered through fine-tuning the pre–trained RFOTNet.

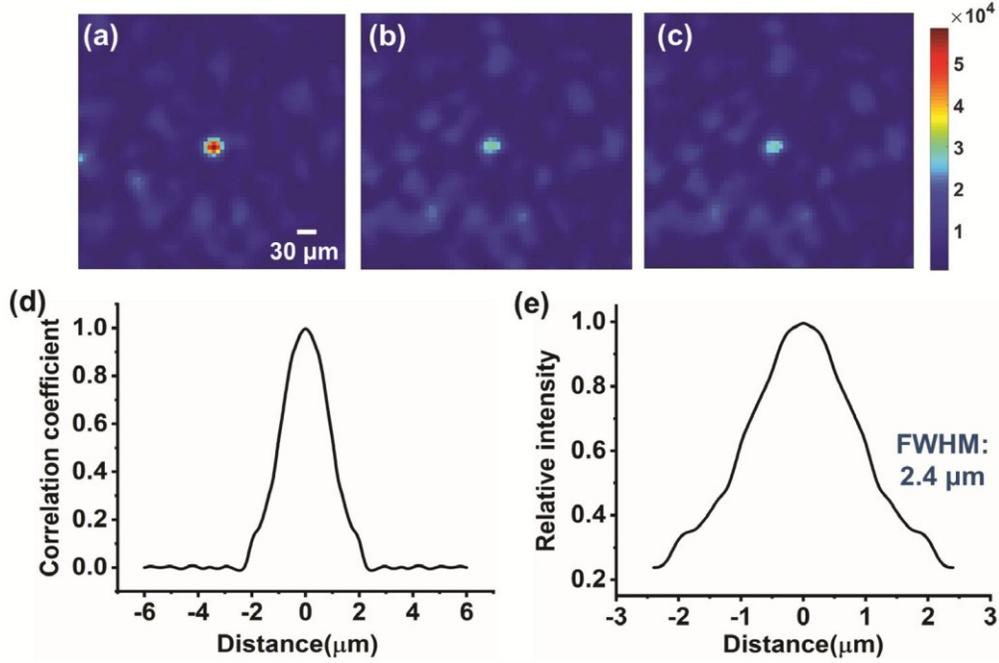

Fig 5. Experimental results of focusing recovery with memory effect. (a) Original focused speckle obtained using the RFOTNet. (b) Speckle pattern after moving the diffuser left by 1μm while the SLM pattern is the same as (a). (c) Speckle pattern after the SLM shifted left accordingly. (d) Correlation between original focus (a) and speckles after medium change (with the same SLM pattern). (e) Focal peak intensity as a function of diffuser translation distance. Note that Figs. 6a-c use the same colormap and scalar bar.

The feasibility of our method has been clearly demonstrated, as shown in Fig. 4. We can also see that after the medium movement, the enhancements $\eta$ is degraded to different extents. Fig. 6 further illustrates how the drop of $\eta$ is affected by the degree of medium movement. The original focus is the same, so the influence from the focusing performance has been eliminated; correlation coefficient serves as the only convincing factor that describes the change of speckle patterns and measures the difficulty of focus recovery. As seen, when the correlation coefficient is larger than 0.72, the enhancements $\eta$ drops to 0.7-0.9 of the original value. In this situation, medium change is mild, and the refocusing is relatively easier. When the correlation coefficient goes lower than 0.5, suggesting a larger medium change, the enhancements $\eta$ drops to less than half of its original value. In this circumstance, recovering the focusing can be difficult. Generally speaking, larger is the medium change, smaller is the correlation coefficient before and after the change, more reduction is the enhancements $\eta$, and more challenging is the focusing recovery.

Next, let's examine how the focusing recovery performance is related with the amount of fine-tuning samples and the degree of medium change. Experiments were conducted five times, with averaged results shown in Fig. 7. Note that the original focal point was the same, and the enhancements $\eta$ recovery percentage is measured by the ratio between the $\eta$ after the RFOTNet fine-tuning and the original $\eta$ before the medium change. 3000, 5000, and 6000 samples were used to fine-tune the pre-trained RFOTNet, and the corresponding results are shown in grey, red, and blue colors, respectively. As seen, when the correlation coefficient is 0.7-0.8 (mild medium perturbation), 3000 samples for fine-tuning is enough to fully restore the original enhancements $\eta$. When the correlation coefficient is reduced to 0.4-0.5 and 0.3, respectively, 5000 and 6000 samples are required accordingly to recover the original performance. If the same amount of tuning samples is provided, the higher is the correlation coefficient, the smaller is the medium change, and the higher is the refocusing quality (as measured by the enhancements $\eta$). Although there are some unavoidable interference and/or noise in experiment, the experimental results agree well with simulation. Compared with existing methods such as iterative optimization and transmission matrix, our method needs less measurements and simpler setup to realize focusing recovery. It is worth noting that under some circumstances, the enhancement $\eta$ recovery percentage can be larger than 1, indicating that the focused speckle achieved by the fine-tuned RFOTNet may even be superior to the original pre-trained RFOTNet. This phenomenon manifests one attractive merit of the proposed method that the fine-tuning of a neural network is not only capable to adapt it to different medium status, it also can compensate the deficiency of the pre-trained network, improving it towards optimal performance.

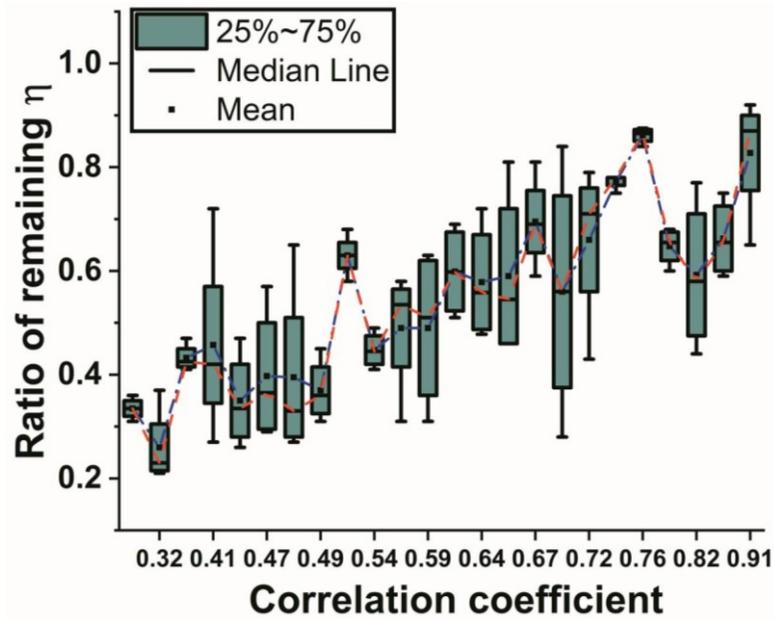

Fig.6. The relationship between the enhancements $\eta$ remaining ratio and the levels of scattering medium change. The original focal point was the same, and the diffuser was moved by various distances to introduce different levels of perturbations. All results are scaled as the original $\eta$ to be 1. Red dashed line illustrates the change of median values while blue dotted line shows the mean values.

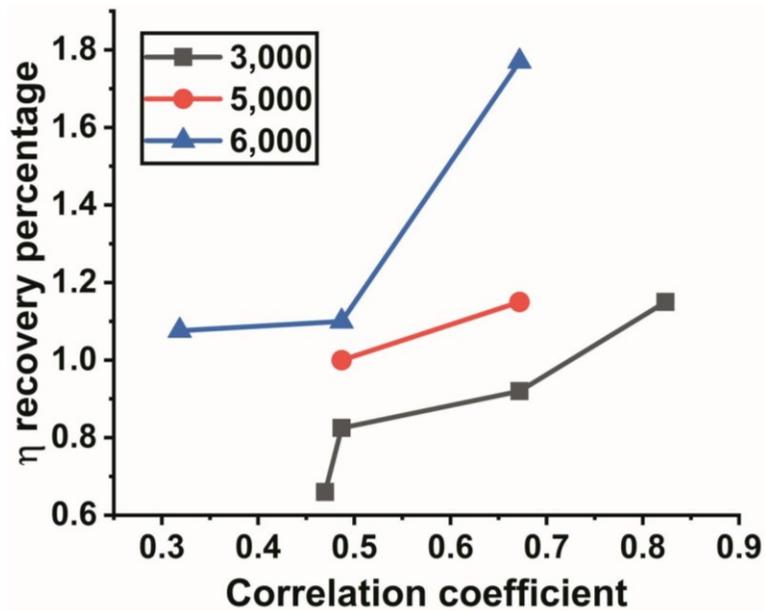

Fig.7. The focusing recovery performance, as measured by the relative enhancements $\eta$, as a function of the amount of samples in fine-tuning and the degree of medium change.

Lastly, since the performance of neural networks highly relies on the training samples, using more samples will definitely lead to better focusing and refocusing, at the expense of spending more time and memory in sample collection and training. So far such trade-off between performance and time cannot be eliminated. Currently, the speed in our study is mainly limited by the slow frame rate of the liquid crystal-based SLM, whose nominal frame rate is up to 60 Hz, but due to the existence of rising and falling response time, in our experiment, the SLM operated at only 16.67 Hz. Digital micromirror device (DMD) can be much faster (the settling time is only 18 μs[58]), but it can only provide binary magnitude modulation for incident optical wavefronts. Thus, the focusing performance is inherently poorer; the enhancements $\eta$ is only 1/5 of that with SLM when the same number of modulating megapixels are actively assigned[59]. In the meanwhile, to speed up the computation, a more powerful GPU and the adoption of FPGA are highly desired. In summary, the proposed method has demonstrated great potential in implementing wavefront shaping-based optical focusing in dynamic media, which could be of great significance to the field.

**Methods**

**Theoretical analysis of implementing deep neural networks for light refocusing through nonstationary scattering media**

The scenario is that a monochromatic optical wave field propagates from the source to a scattering layer, and the transmitted scattered light is collected by a camera. After scattering, the optical field at the receiving plane $r_c$ (e.g. the camera plane in Fig. 1a) is given using the Fresnel-Kirchhoff diffraction formula[60], where the effects of absorption is neglected,

$$E(\boldsymbol{r}_c) = \iint d^2\boldsymbol{r}_b g(r_c - r_b)\frac{i}{\lambda}E(r_b) = \iint d^2 r_b g(r_c - r_b)\frac{i}{\lambda}\left(\frac{i}{\lambda}\iint d^2 r_a G(r_b, r_a)E(r_a)\right), \quad (1)$$

where $r_a$ is the source plane (e.g., the SLM plane in Fig. 1a), $r_b$ is the back surface of the scattering layer. $E(r_a)$, $E(r_b)$, and $E(r_c)$ are optical fields at $r_a$, $r_b$ and $r_c$, respectively. $G(r_b, r_a)$ is the unknown Green function for light traveling from $r_a$ to $r_b$, while $g(r_c - r_b)$ is the free space Green function for light traveling from $r_b$ to $r_c$[60]. Regular cameras only record the light intensity distribution of the speckle patterns on $r_c$, thus

$$I(r_c) = |E(r_c)|^2. \quad (2)$$

Based on the above forward scattering model, to precisely compute the required incident complex optical field $E(r_a)$ with which light is focused through the scattering medium, the inverse scattering model has to be obtained based on the recorded transmitted light intensity distribution $I(r_c)$. From Eqs. (1) and (2), we can see that the inverse scattering problem is nonlinear and ill-posed. These properties prohibit the adoption of direct inversion methods; iterative optimization with regularization is necessary[51] to encounter this problem.

In order to fit the general regularized formulation and solve the specific light scattering problem, here we combine Eqs. (1) and (2) into a discretized form

$$I_c = \left|\sum_{n=1}^{N} t_{cn}(r_a)E_n(r_a)\right|^2, \quad (3)$$

where $t_{cn}(r_a)$ is a complex transmission coefficient describing light propagation from the $n^{th}$ optical mode on plane $r_a$ to the receiving plane $r_c$. Therefore, the general Tikhonov regularized formulation can be given as

$$\hat{f}_\alpha = \arg\min_{f} \|I_c - Hf\|^2 + \alpha \|f\|^2, \tag{4}$$

where $H$ is the forward scattering model relating the measured speckle patterns $I_c$ and light source $f$, $I_c = Hf$, and α is the regulation parameter. In our wavefront shaping, $f$ is a matrix $E^a$ that consists of $N$ different optical modes, $E^a = [E_1(r_a)\ E_2(r_a)\ \cdots E_N(r_a)]$.

A lot of iterative methods, such as distorted Born iterative method[61], subspace optimization method (SOM)[62], and iterative shrinkage and thresholding algorithm (ISTA)[63], have been reported to solve Eq. (4). Among them, Yuan et al. prove the existence and efficacy of the inverse scattering function in light scattering problem[64], Xu et al. confirm that CNNs can be used to solve the inverse scattering problem thanks to the kernel separability[31], and Jin et al. construct a model revealing the intrinsic relationship between the CNN and the iterative optimization methods for shift-invariant systems[65]. Note that from Eq. (1), due to multiple scattering, the whole system is no longer shift-invariant. As CNN has the capability to learn the degree of shift variance and partially compensate it[44], CNN has the potential to resolve inverse scattering problems.

As regard to Eq. (4), most conventional iterative algorithms rely on the building block model[65]:

$$p^{m+1} = A_\theta \left[ \frac{1}{L} W * H * I_c + \left( I - \frac{1}{L} W * H * HW \right) p^m \right] \tag{5}$$

where the desired estimation $\hat{E}^a = Wp$, in which $W$ is a convolutional transformation and $p$ is the transformation coefficients. $L$ is the Lipschitz constant, $L \le eig(W * H * HW)$. From Eq. (5), the iterative optimization process can be regarded as a sequence of linear filtering by kernel $I - (1/L)W * H * HW$ and bias $(1/L)W * H * I_c$, followed by a point-wise nonlinear operation $A_\theta$ by value $\theta$. This is actually quite similar to a typical structure of CNN[65]. As the abovementioned iterative methods can effectively (albeit time consuming) solve the inverse scattering problem, it is natural to hypothesize that CNNs can be implement to tackle the problem.

As a pioneering work, in this study we directly apply a deep CNN to model the inverse scattering function $H^{-1}$ and establish the relationship between the speckle intensity distribution $I_c$ and its corresponding incident optical wavefront $\hat{E}^a$. The process can be formulated as a supervised learning problem[53]. Note that we will not compare the parameters in iterative algorithms with the CNN weights. When the scattering medium is subject to perturbations, both $H$ and $H^{-1}$ change accordingly. The speckle correlation theory in random media suggests that, when the configurations of the scatterers are changed randomly, the scattering media before and after moderate change are correlated[66], which can be shown as below

$$\begin{aligned}
C_{aba'b'} &= C^{(1)}_{aba'b'} + C^{(2)}_{aba'b'} + C^{(3)}_{aba'b'} \\
&= D_1 \langle T_{ab} \rangle \langle T_{a'b'} \rangle \delta_{\Delta k_a, \Delta k_b} F_1(\Delta k_a L) \\
&\quad + D_2 g^{-1} \langle T_{ab} \rangle \langle T_{a'b'} \rangle \delta_{\Delta k_a, \Delta k_b} \left[ F_2(\Delta k_a L) + F_2(\Delta k_b L) \right] \\
&\quad + D_3 g^{-2} \langle T_{ab} \rangle \langle T_{a'b'} \rangle
\end{aligned} \tag{6}$$

Here, $T_{ab}$ is the transmission coefficient between the incident mode $a$ and transmitted mode $b$ (similar for $T_{a'b'}$), $D_1, D_2, D_3$ are constants, $\Delta k_a \equiv |k_a - k_{a'}|$ (similar for $\Delta k_b$), where $k_a$ and $k_b$ are the incident and transmitted wave vectors, respectively. $F_1(x)$ and $F_2(x)$ are form-factor function. $C^{(1)}_{aba'b'}$, $C^{(2)}_{aba'b'}$, and $C^{(3)}_{aba'b'}$ contribute to the short-range correlation, long-range correlation, and infinite-range correlation, respectively[67]. For most scattering media, the magnitude of $C^{(1)}_{aba'b'}$, $C^{(2)}_{aba'b'}$, and $C^{(3)}_{aba'b'}$ decrease in sequence. After perturbation, the medium is changed to some extent, and the correlation before and after the change is governed by $C_{aba'b'}$. Based on the correlation, the inverse scattering function $H_2^{-1}$ after the change can be deduced from $H^{-1}$ as long as the CNN has learnt the difference between these two states, calling for the necessity of adopting the fine-tuning of CNN using samples collected from the changed state. Thanks again to the correlation, the amount of fine-tuning samples can be much less than the samples used for training the CNN from random initialization, which contributes meaningfully to decreasing the time spent in re-optimization. From Eq. (6), it is obvious that the larger the change, the smaller the correlation coefficient $C_{aba'b'}$, and more samples will be needed for fine-tuning. After fine tuning, the new CNN is able to adapt to the new relationship between the speckles and incident light wavefronts for the perturbed medium.

**Experiment setup**

Experiment setup is illustrated in Fig. 3. Light emitting from a He-Ne CW laser (633nm, Melles Griot) is expanded by a telescope by 4.3 times. Then a half-wave plate and a polarizer are followed to adjust the polarization of the incident light to be parallel to the long axis of a SLM (X13138-01, Hamamatsu). The light wavefront is modulated by the SLM, after which the light passes through two successive lenses and is focused onto the surface of a diffuser (120 Grit, Edmund ground glass diffuser,) by an objective lens (TU Plan Fluor 50X/0.80, Nikon). The light undergoes multiple scatterings inside the diffuser, then the scattered light is collected by another objective lens (TU Plan Fluor 20X/0.45, Nikon) placed behind the diffuser. Finally the speckles are recorded via a camera (Zyla s4.2, Andor). For the diffuser used, the mean free path $l$ is around 18 μm, while the thickness of the scattering surface $L$ is 0.25mm ~ 0.5 mm, $L \gg l$.

**RFOTNet structure**

U-Net is widely used to solve the image reconstruction problems through scattering media[51,65,68]. Image features are extracted and abstracted in contracting paths, which significantly enlarges the effective receptive field of the network and improves the output prediction[51]. Then expansion paths upsample feature maps back to the original image resolution[69]. Nevertheless, image reconstruction concerns more about the quality of the restored images, while it does not pose high demands on time. As for wavefront shaping, time is critical due to the limitation of decorrelation time. Thus a simpler while efficient network specifically targeting on learning the information of scattering processes and modelling inverse functions is preferred in order to speed up SLM patterns computation.

The structure of the proposed RFOTNet is shown in Fig. 1(b). This RFOTNet has three inputs and one output. Input 1 is the speckle pattern recorded by a camera, while the corresponding phase pattern of the light entering the scattering medium is noted as Input 2. Since phase patterns of light are usually adjusted by a spatial light modulator (SLM), they are represented by SLM patterns. Input 3 is the targeted speckle pattern, while RFOTNet output is the SLM pattern needed in order to get Input 3 through the scattering medium. Each time the scattering medium is changed, the RFOTNet has to be adjusted to model the new medium. In transfer learning, generally only the last few layers, rather than the whole neural networks, are fine-tuned[70-73] as the last layers are task specific while the earlier ones are modality specific[74]. Information learnt by earlier layers can be shared among all inverse scattering problems, while the last few layers are customized for adapting to specialized conditions. Therefore, in our experiment, only the last fully-connected layer (FC2) in the RFOTNet is fine-tuned while all the other layers are frozen. By doing so, it also saves time and computational resources. The three convolutional layers, Conv1, Conv2, and Cov3, extract image features from Input 1, then these features are flattened to a 1D array to concatenate with Input 2, which has also been flattened. The combination serves as the input to the first fully-connected layer (FC1), followed by a dropout layer. The outputs of FC1 concatenate with the image features extracted from Input 3. The final fully-connected layer (FC2) predicts the SLM pattern needed for Input 3. Conv1, Conv2, and Cov3 consist of 16, 32, and 48 filters, respectively, and the filter size of each layer is $7 \times 7$, $5 \times 5$, and $3 \times 3$ with stride setting as $3 \times 3$, $2 \times 2$, and 1 × 1, respectively. Conv1 and Conv4, Conv2 and Conv5, and Conv3 and Conv6 have the same structure, respectively. The number of neurons in FC1 is 512 with a dropout rate set to 0.5. The number of neurons in FC2 is the same as the size of the SLM patterns. Kernel initializers of all layers are set as glorot normal. Mean squared error is employed as the loss function. Adam is used as the optimizer with alpha, beta1, beta2, and epsilon set as 0.0005, 0,9, 0.99, and 0.0001, respectively. The activation function of all layers are tanh, except for the last output layer whose activation function is sigmoid. Tensorflow Keras library is used to construct the model, and the GPU used in computation is NVIDIA GEFORCE GTX 980.

## Acknowledgement

The work has been supported by the A*STAR SERC AME Program: Nanoantenna Spatial Light Modulators for Next Generation Display Technologies (Grant No. A18A7b0058), the National Natural Science Foundation of China (No. 81671726, No. 81627805, and No. 81930048), the Hong Kong Research Grant Council (No. 25204416), the Hong Kong Innovation and Technology Commission (no. ITS/022/18), and the Shenzhen Science and Technology Innovation Commission (No. JCYJ20170818104421564).

## Conflict of interest

The authors declare no conflict of interest.

## Authors contributions

Y.L. and H.L. worked on the simulation. S.Y., Y.L., and H.L. contributed to the experiments. P.L. and Y.Z. conceived and supervised the research. All authors contributed to discussing, editing, and revising the paper.

**Supplementary information** accompanies the manuscript.

## References

1      Bertolotti, J. *et al.* Non-invasive imaging through opaque scattering layers. *Nature* **491**, 232 (2012).
2      Lai, P., Wang, L., Tay, J. W. & Wang, L. V. Photoacoustically guided wavefront shaping for enhanced optical focusing in scattering media. *Nat Photonics* **9**, 126-132, doi:10.1038/nphoton.2014.322 (2015).
3      Xu, X., Liu, H. & Wang, L. V. Time-reversed ultrasonically encoded optical focusing into scattering media. *Nat Photonics* **5**, 154 (2011).
4      Lai, P., Xu, X., Liu, H. & Wang, L. V. Time-reversed ultrasonically encoded optical focusing in biological tissue. *J Biomed Opt* **17**, 030506, doi:10.1117/1.JBO.17.3.030506 (2012).
5      Judkewitz, B., Wang, Y. M., Horstmeyer, R., Mathy, A. & Yang, C. Speckle-scale focusing in the diffusive regime with time-reversal of variance-encoded light (TROVE). *Nat Photonics* **7**, 300-305, doi:10.1038/nphoton.2013.31 (2013).
6      Lai, P., Xu, X., Liu, H., Suzuki, Y. & Wang, L. V. Reflection-mode time-reversed ultrasonically encoded optical focusing into turbid media. *Journal of Biomedical Optics* **16**, 080505 (2011).
7      Wang, Y. M., Judkewitz, B., DiMarzio, C. A. & Yang, C. Deep-tissue focal fluorescence imaging with digitally time-reversed ultrasound-encoded light. *Nature communications* **3**, 928 (2012).
8      Yaqoob, Z., Psaltis, D., Feld, M. S. & Yang, C. Optical phase conjugation for turbidity suppression in biological samples. *Nature photonics* **2**, 110 (2008).
9      Vellekoop, I. M. & Mosk, A. P. Focusing coherent light through opaque strongly scattering media. *Opt Lett* **32**, 2309-2311 (2007).
10      Popoff, S. M. *et al.* Measuring the transmission matrix in optics: an approach to the study and control of light propagation in disordered media. *Phys Rev Lett* **104**, 100601, doi:10.1103/PhysRevLett.104.100601 (2010).
11      Katz, O., Small, E. & Silberberg, Y. Looking around corners and through thin turbid layers in real time with scattered incoherent light. *Nature photonics* **6**, 549 (2012).
12      Yu, H. *et al.* Measuring large optical transmission matrices of disordered media. *Physical review letters* **111**, 153902 (2013).
13      Chaigne, T. *et al.* Controlling light in scattering media non-invasively using the photoacoustic transmission matrix. *Nature Photonics* **8**, 58 (2014).
14      Takasaki, K. T. & Fleischer, J. W. Phase-space measurement for depth-resolved memory-effect imaging. *Optics express* **22**, 31426-31433 (2014).

15. Katz, O., Heidmann, P., Fink, M. & Gigan, S. Non-invasive single-shot imaging through scattering layers and around corners via speckle correlations. *Nature photonics* **8**, 784 (2014).
16. Katz, O., Small, E., Guan, Y. & Silberberg, Y. Noninvasive nonlinear focusing and imaging through strongly scattering turbid layers. *Optica* **1**, 170-174 (2014).
17. Edrei, E. & Scarcelli, G. Memory-effect based deconvolution microscopy for super-resolution imaging through scattering media. *Scientific reports* **6**, 33558 (2016).
18. Thompson, J., Hokr, B. & Yakovlev, V. Optimization of focusing through scattering media using the continuous sequential algorithm. *Journal of modern optics* **63**, 80-84 (2016).
19. Conkey, D. B., Brown, A. N., Caravaca-Aguirre, A. M. & Piestun, R. Genetic algorithm optimization for focusing through turbid media in noisy environments. *Optics express* **20**, 4840-4849 (2012).
20. Dong, B. & Booth, M. J. Wavefront control in adaptive microscopy using Shack-Hartmann sensors with arbitrarily shaped pupils. *Optics express* **26**, 1655-1669 (2018).
21. Rahman, S. A. & Booth, M. J. Direct wavefront sensing in adaptive optical microscopy using backscattered light. *Applied optics* **52**, 5523-5532 (2013).
22. Schwertner, M., Booth, M. & Wilson, T. Wavefront sensing based on rotated lateral shearing interferometry. *Optics Communications* **281**, 210-216 (2008).
23. Park, J.-H., Yu, Z., Lee, K., Lai, P. & Park, Y. Perspective: Wavefront shaping techniques for controlling multiple light scattering in biological tissues: Toward in vivo applications. *APL photonics* **3**, 100901 (2018).
24. Bossy, E. & Gigan, S. Photoacoustics with coherent light. *Photoacoustics* **4**, 22-35, doi:10.1016/j.pacs.2016.01.003 (2016).
25. Benjamin Judkewitz, R. H., Ivo M. Vellekoop, Ioannis N. Papadopoulos and Changhuei Yang. Translation correlations in anisotropically scattering media. *Nat Phys* **11**, 684-689, doi:10.1038/NPHYS3373 (2015).
26. LeCun, Y., Bengio, Y. & Hinton, G. Deep learning. *nature* **521**, 436 (2015).
27. Xie, J., Xu, L. & Chen, E. in *Advances in neural information processing systems*. 341-349.
28. Burger, H. C., Schuler, C. J. & Harmeling, S. in *2012 IEEE conference on computer vision and pattern recognition*. 2392-2399 (IEEE).
29. Agostinelli, F., Anderson, M. R. & Lee, H. in *Advances in Neural Information Processing Systems*. 1493-1501.
30. Schuler, C. J., Christopher Burger, H., Harmeling, S. & Scholkopf, B. in *Proceedings of the IEEE Conference on Computer Vision and Pattern Recognition*. 1067-1074.
31. Xu, L., Ren, J. S., Liu, C. & Jia, J. in *Advances in Neural Information Processing Systems*. 1790-1798.
32. Lucas, A., Iliadis, M., Molina, R. & Katsaggelos, A. K. Using deep neural networks for inverse problems in imaging: beyond analytical methods. *IEEE Signal Processing Magazine* **35**, 20-36 (2018).
33. Mousavi, A., Patel, A. B. & Baraniuk, R. G. in *2015 53rd Annual Allerton Conference on Communication, Control, and Computing (Allerton)*. 1336-1343 (IEEE).
34. Barbastathis, G., Ozcan, A. & Situ, G. On the use of deep learning for computational imaging. *Optica* **6**, 921-943 (2019).
35. Sinha, A., Lee, J., Li, S. & Barbastathis, G. Lensless computational imaging through deep learning. *Optica* **4**, 1117-1125 (2017).
36. Deng, M., Li, S. & Barbastathis, G. Learning to synthesize: splitting and recombining low and high spatial frequencies for image recovery. *arXiv preprint arXiv:1811.07945* (2018).
37. Waller, L. & Tian, L. Computational imaging: Machine learning for 3D microscopy. *Nature* **523**, 416 (2015).
38. Goy, A., Arthur, K., Li, S. & Barbastathis, G. Low photon count phase retrieval using deep learning. *Physical review letters* **121**, 243902 (2018).
39. McCann, M. T., Jin, K. H. & Unser, M. Convolutional neural networks for inverse problems in imaging: A review. *IEEE Signal Processing Magazine* **34**, 85-95 (2017).
40. Cui, Z., Chang, H., Shan, S., Zhong, B. & Chen, X. in *European Conference on Computer Vision*. 49-64 (Springer).
41. Dong, C., Loy, C. C., He, K. & Tang, X. Image super-resolution using deep convolutional networks. *IEEE transactions on pattern analysis and machine intelligence* **38**, 295-307 (2015).
42. Turpin, A., Vishniakou, I. & d Seelig, J. Light scattering control in transmission and reflection with neural networks. *Optics express* **26**, 30911-30929 (2018).
43. Li, Y., Xue, Y. & Tian, L. Deep speckle correlation: a deep learning approach toward scalable imaging through scattering media. *Optica* **5**, 1181-1190 (2018).
44. Li, S., Deng, M., Lee, J., Sinha, A. & Barbastathis, G. Imaging through glass diffusers using densely connected convolutional networks. *Optica* **5**, 803-813 (2018).
45. Rahmani, B., Loterie, D., Konstantinou, G., Psaltis, D. & Moser, C. Multimode optical fiber transmission with a deep learning network. *Light: Science & Applications* **7**, 69 (2018).
46. Luo, Q., Newman, J. A. & Webb, K. J. Motion-based coherent optical imaging in heavily scattering random media. *Optics Letters* **44**, 2716-2719 (2019).
47. Porat, A. *et al.* Widefield lensless imaging through a fiber bundle via speckle correlations. *Optics express* **24**, 16835-16855 (2016).
48. Yilmaz, H. *et al.* Speckle correlation resolution enhancement of wide-field fluorescence imaging. *Optica* **2**, 424-429 (2015).
49. Sun, Y., Shi, J., Sun, L., Fan, J. & Zeng, G. Image reconstruction through dynamic scattering media based on deep learning. *Optics express* **27**, 16032-16046 (2019).
50. Nissilä, I., Noponen, T., Heino, J., Kajava, T. & Katila, T. (Springer, 2005).

51  Wei, Z. & Chen, X. Deep-learning schemes for full-wave nonlinear inverse scattering problems. *IEEE Transactions on Geoscience and Remote Sensing* **57**, 1849-1860 (2018).
52  Sinha, A., Lee, J., Li, S. & Barbastathis, G. in *Digital Holography and Three-Dimensional Imaging.* W1A.3 (Optical Society of America).
53  Horisaki, R., Takagi, R. & Tanida, J. Learning-based imaging through scattering media. *Opt Express* **24**, 13738-13743, doi:10.1364/OE.24.013738 (2016).
54  Feng, S., Kane, C., Lee, P. A. & Stone, A. D. Correlations and fluctuations of coherent wave transmission through disordered media. *Physical review letters* **61**, 834 (1988).
55  Li, J. & Genack, A. Correlation in laser speckle. *Physical Review E* **49**, 4530 (1994).
56  Schott, S., Bertolotti, J., Léger, J.-F., Bourdieu, L. & Gigan, S. Characterization of the angular memory effect of scattered light in biological tissues. *Optics express* **23**, 13505-13516 (2015).
57  Fayyaz, Z., Mohammadian, N. & Avanaki, M. R. in *Photons Plus Ultrasound: Imaging and Sensing 2018.* 104946I (International Society for Optics and Photonics).
58  Dudley, D., Duncan, W. M. & Slaughter, J. in *MOEMS display and imaging systems.* 14-26 (International Society for Optics and Photonics).
59  Yu, Z., Li, H. & Lai, P. Wavefront shaping and its application to enhance photoacoustic imaging. *Applied Sciences* **7**, 1320 (2017).
60  Vellekoop, I. M. Controlling the propagation of light in disordered scattering media. *arXiv preprint arXiv:0807.1087* (2008).
61  Chew, W. C. & Wang, Y.-M. Reconstruction of two-dimensional permittivity distribution using the distorted Born iterative method. *IEEE transactions on medical imaging* **9**, 218-225 (1990).
62  Chen, X. Subspace-based optimization method for solving inverse-scattering problems. *IEEE Transactions on Geoscience and Remote Sensing* **48**, 42-49 (2009).
63  Kamilov, U. S. & Mansour, H. Learning optimal nonlinearities for iterative thresholding algorithms. *IEEE Signal Processing Letters* **23**, 747-751 (2016).
64  Yuan, Z. & Wang, H. Multiple Scattering Media Imaging via End-to-End Neural Network. *arXiv preprint arXiv:1806.09968* (2018).
65  Jin, K. H., McCann, M. T., Froustey, E. & Unser, M. Deep convolutional neural network for inverse problems in imaging. *IEEE Transactions on Image Processing* **26**, 4509-4522 (2017).
66  Feng, S., Kane, C., Lee, P. A. & Stone, A. D. Correlations and fluctuations of coherent wave transmission through disordered media. *Phys Rev Lett* **61**, 834-837, doi:10.1103/PhysRevLett.61.834 (1988).
67  Sebbah, P. *Waves and imaging through complex media.* (Springer Science & Business Media, 2001).
68  Han, Y. & Ye, J. C. Framing U-Net via deep convolutional framelets: Application to sparse-view CT. *IEEE transactions on medical imaging* **37**, 1418-1429 (2018).
69  Ronneberger, O., Fischer, P. & Brox, T. in *International Conference on Medical image computing and computer-assisted intervention.* 234-241 (Springer).
70  Yosinski, J., Clune, J., Bengio, Y. & Lipson, H. in *Advances in neural information processing systems.* 3320-3328.
71  Shin, H.-C. *et al.* Deep convolutional neural networks for computer-aided detection: CNN architectures, dataset characteristics and transfer learning. *IEEE transactions on medical imaging* **35**, 1285-1298 (2016).
72  Christodoulidis, S., Anthimopoulos, M., Ebner, L., Christe, A. & Mougiakakou, S. Multisource transfer learning with convolutional neural networks for lung pattern analysis. *IEEE journal of biomedical and health informatics* **21**, 76-84 (2016).
73  Girshick, R., Donahue, J., Darrell, T. & Malik, J. Region-based convolutional networks for accurate object detection and segmentation. *IEEE transactions on pattern analysis and machine intelligence* **38**, 142-158 (2015).
74  Castrejon, L., Aytar, Y., Vondrick, C., Pirsiavash, H. & Torralba, A. in *Proceedings of the IEEE Conference on Computer Vision and Pattern Recognition.* 2940-2949.

**Supplementary Information**

Simulation was conducted to compare the performance of the proposed RFOTNet and a conventional CNN in light focusing and refocusing through a nonstationary scattering medium. The structure of RFOTNet is shown in Fig. 1(b), while the structure of the constructed conventional CNN is shown in Fig. S1. The CNN consists of three convolutional layers and two fully connected layers. The input of the CNN is speckle patterns while output is their corresponding SLM patterns. All convolutional (Conv 1, Conv 2, and Conv 3) and fully-connected layers (FC1 and FC2) share the same structure with the convolutional (Conv 1, Conv 2, and Conv 3) and fully-connected layers (FC1 and FC2) in RFOTNet, respectively. Training and fine-tuning settings are also the same. First of all, 10,000 samples were generated using a transmission matrix to train the both networks. After training, the same focused speckle was sent to the pretrained RFOTNet and CNN, and the focusing results with the SLM pattern predicted by them are shown in Fig. S2 (c) and (e), respectively. Note that Fig. S2c-f use the same color bar. As shown in Fig. S2 (a), RFOTNet demonstrated much higher converging speed than the conventional CNN during training, and the focusing result (Fig. S2c) is also much better. Then the environmental perturbation was modelled by adding another matrix to the original transmission matrix. After perturbation, the optical focusing was degraded. Then 3,000 fine-tuning samples were generated based on the new transmission matrix. During the fine-tuning process, RFOTNet still converged faster (Fig. S2b). Focusing recovery results by RFOTNet and CNN are shown in Fig. S2 (d) and (f), respectively. RFOTNet not only demonstrates superior ability in establishing the transformation from speckles to SLM patterns, it is also more powerful in adapting to the new medium state. In RFOTNet, Input 1 and Input 2 provide rich information about the scattering medium to facilitate the establishment of a more precise mapping from Input 3 to Output. Moreover, during fine-tuning, RFOTNet is better at discovering the difference before and after perturbation, and adjusting the weights to model the new status.

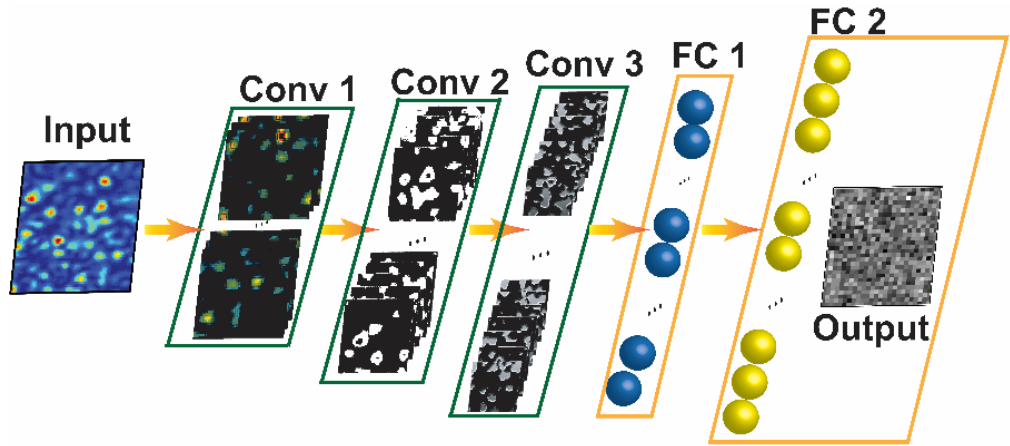

Fig. S1. Structure of a conventional CNN. Input is speckle patterns and output is corresponding SLM patterns. The structure of all the convolutional layers and fully-connected layers in the CNN are the same as those in RFOTNet.

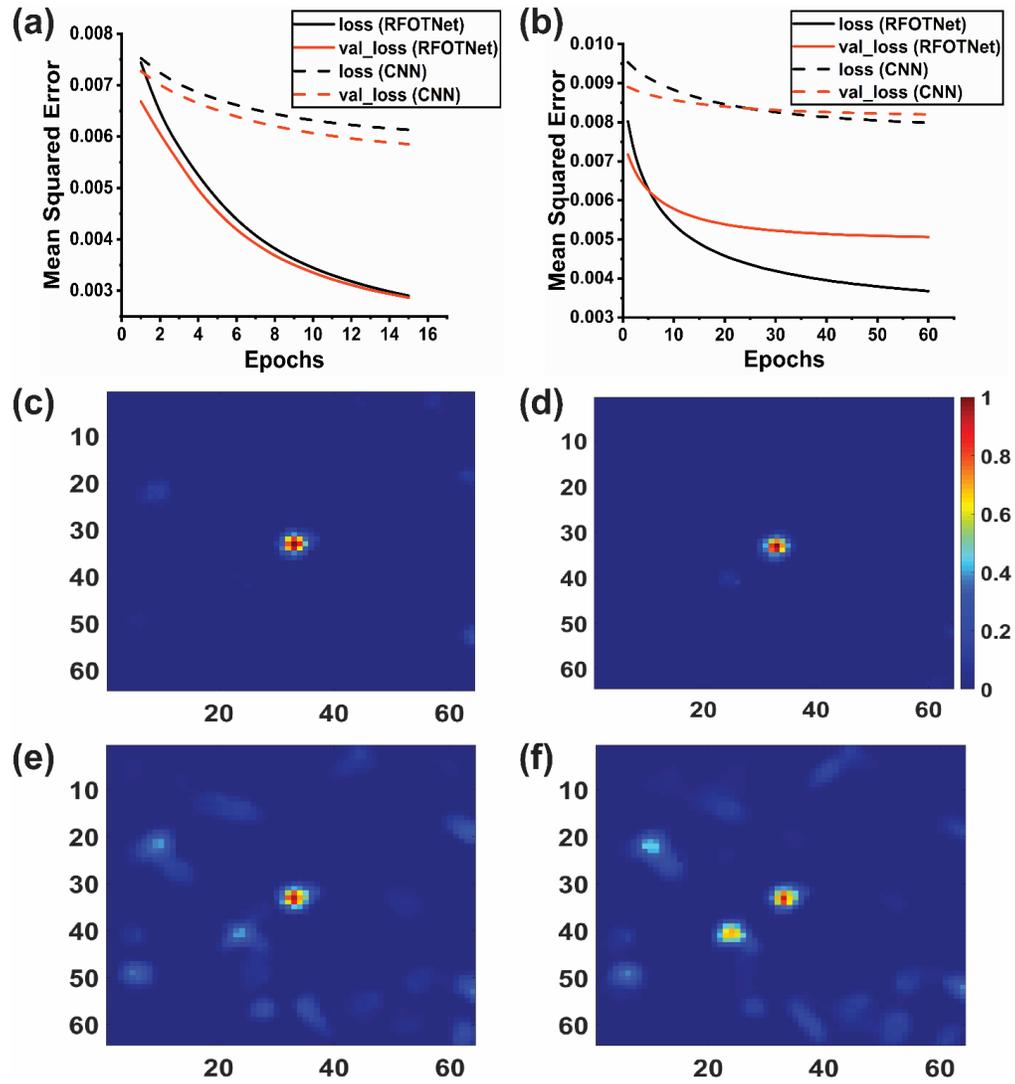

Fig. S2. Simulation results of RFOTNet and conventional CNN in light focusing and refocusing. (a) Training performance of RFOTNet and CNN. Straight lines indicate the results of RFOTNet, while dashed lines show the results of CNN. Black lines show the loss values (mean-squared-error) during training, and red lines show the validation loss values (mean-squared-error). (b) The performance of RFOTNet and CNN during fine-tuning. (c) Focused speckle obtained using RFOTNet after training. (d) Focusing recovery results with fine-tuned RFOTNet after medium change. (e) Focused speckle obtained using CNN after training. (f) Focusing recovery results with fine-tuned CNN after medium change. Note that (c)-(f) use the same color bar.